\documentclass[11pt,a4paper]{article}
\usepackage{amsmath, amssymb, natbib, graphicx}
\usepackage{lineno,setspace}

\title{Dilatancy-induced fluid pressure drop during dynamic rupture: Direct experimental evidence and consequences for earthquake dynamics}
\author{Nicolas \textsc{Brantut}\\Department of Earth Sciences\\University College London\\London, WC1E6BS, UK}
\date{\ }

\begin{document}

\maketitle

\begin{abstract}
Fluid pressure and flow in the crust is a key parameter controlling earthquake physics. Since earthquake slip is linked to spatio-temporal localisation of deformation, it is expected that the localised fluid pressure around the fault plane could potentially impact the dynamic strength of the slipping fault zone. Coseismic fluid pressure drops have been inferred from field studies, notably in gold deposits which are thought to be formed by this process, but reliable quantitative predictions are still lacking.  Here, experimental results are presented where local on- and off-fault fluid pressure variations were measured \textit{in situ} during dynamic rock fracture and frictional slip under upper crustal stress conditions. During the main rupture, the on-fault fluid pressure dropped rapidly to zero, indicating partial vaporisation and/or degassing. Further deformation produced stick-slip events systematically associated with near-instantaneous drops in fluid pressure, providing  direct experimental support of the concept of ``seismic suction pump''. In situ fluid volume and wave speed measurements together with microstructural investigations show that dilatancy is the process driving fluid pressure drops during rupture and slip. Extrapolation of the laboratory results indicate that dilatancy-induced  fluid pressure drops might be a widespread phenomenon in the crust, counteracting thermal pressurisation as a weakening mechanisms in freshly fractured rock.
\end{abstract}

\section{Introduction}

Pore pressure exerts a direct influence on the strength and stability of faults, but its evolution in time and space is strongly coupled to deformation and faulting processes. Under upper crustal conditions, the formation of macroscopic shear faults is accompanied with dilatancy, i.e., an increase in porosity due to the growth and coalescence of tensional microcracks \citep[e.g.,][]{brace66b}. Under undrained or partially drained conditions, dilatancy is known to produce significant drops in pore fluid pressure \citep[e.g.,][]{brace68}, which has a number of key consequences for fault strength and fluid flow in the crust: a dilatancy-induced decrease in fluid pressure, through the principle of effective stress, tends to strengthen faults \citep[e.g.,][]{rice75b,martin80}, stabilise slip \citep[e.g.,][]{rudnicki88,segall95,segall10}, and may be responsible for episodic hydrothermal circulation in the upper crust \citep{zoback75,sibson75b,sibson94}. Geological records of gold deposits in quartz veins indicate that sudden drops in fluid pressure, probably coseismic, trigger the precipitation of minerals and metals  \citep[e.g.,][]{sibson87, wilkinson96, parry98, cox99, weatherley13}. Fault strengthening due to dilatancy at the onset of seismic slip may also counteract fault weakening due to thermal pressurisation of pore fluids, notably by decreasing the initial pore pressure at the initiation of slip, within the rupture process zone \citep{rice06,rempel06}. Dilatancy can therefore facilitate the occurrence of frictional melting \citep[e.g.,][]{brantut18c}, which has major implications for the dynamics of earthquake slip and energy budget.

While the process of dilatancy is well understood qualitatively, reliable quantitative predictions on its effect on pore pressure remain challenging to produce. Dilatancy has been documented experimentally as a bulk phenomenon due to diffuse microcrack opening prior to strain localisation and faulting \citep[][Chap. 5]{paterson05}. However, the mechanics of faults is also controlled by the dilation occurring within the fault itself. The highly localised nature of rock fracture and fault slip, both in space and time, render the use of averaged, bulk properties (namely, porosity change and material compressibility) inadequate for accurate predictions, which are sensitive to relatively minor variations in poroelastic properties \citep[e.g.,][]{brantut18c}.

Fault zone pore volume change has been measured during slip on synthetic gouge-filled faults \citep[e.g.,][]{morrow89,marone90,samuelson09,faulkner18}, and significant efforts have been devoted to measure dilatancy/compaction during rapid slip events \citep[e.g.,][]{ferri10,violay15}. While synthetic gouge experiments have been instrumental in the development of our understanding of fault zone dilatancy \citep[e.g.,][]{segall95}, they can be challenging to extrapolate to dynamic conditions during earthquakes and to complex fault materials and geometries. In general, volume changes in synthetic gouge are expected to depend significantly on initial consolidation state and the degree of cementation of the gouge. In addition, gouge loss during shear experiments also impacts the accuracy of volumetric measurements for large slip displacements. Finally, gouge experiments are typically performed between planar forcing walls, which masks potential dilatant effects due to natural fault geometry and roughness.

Although the changes in pore volume during deformation and fault slip can be constrained experimentally during friction tests, the impact of dilatancy (or compaction) on pore pressure is often assessed indirectly by modelling the pore pressure evolution using estimates of pore space compressibility and hydraulic diffusivity. Such predictions of pore pressure change are difficult to test directly (although their consistency can be checked through their effect on overall strength evolution, see \citet{faulkner18}), and the impact of porosity change on pore pressure during failure remains mostly unknown.

In order to assess quantitatively the role of dilatancy during rock fracture and seismic slip, experimental measurements of fluid volume or fluid pressure change \emph{within} the fault zone at the inception of slip are needed. This paper reports results from a new experimental methodology developed specifically to obtain direct measurements of on- and near-fault fluid pressure during triaxial rock rupture experiments. These new measurements show that near-fault dilatancy produces a dramatic drop of fluid pressure during rupture of intact crystalline rock, such that the fluid can locally vaporise or degas due to decompression while the fluid pressure a few centimetres away from the fault remains constant. Subsequent stick-slip events on the newly created fault also produce fast drops in fluid pressure. Taken together, these new measurements provide unique constraints on the dilatancy-induced fluid pressure variations occurring during crustal earthquakes in newly formed faults.

\section{Methods}

\begin{figure}
  \centering
  \includegraphics{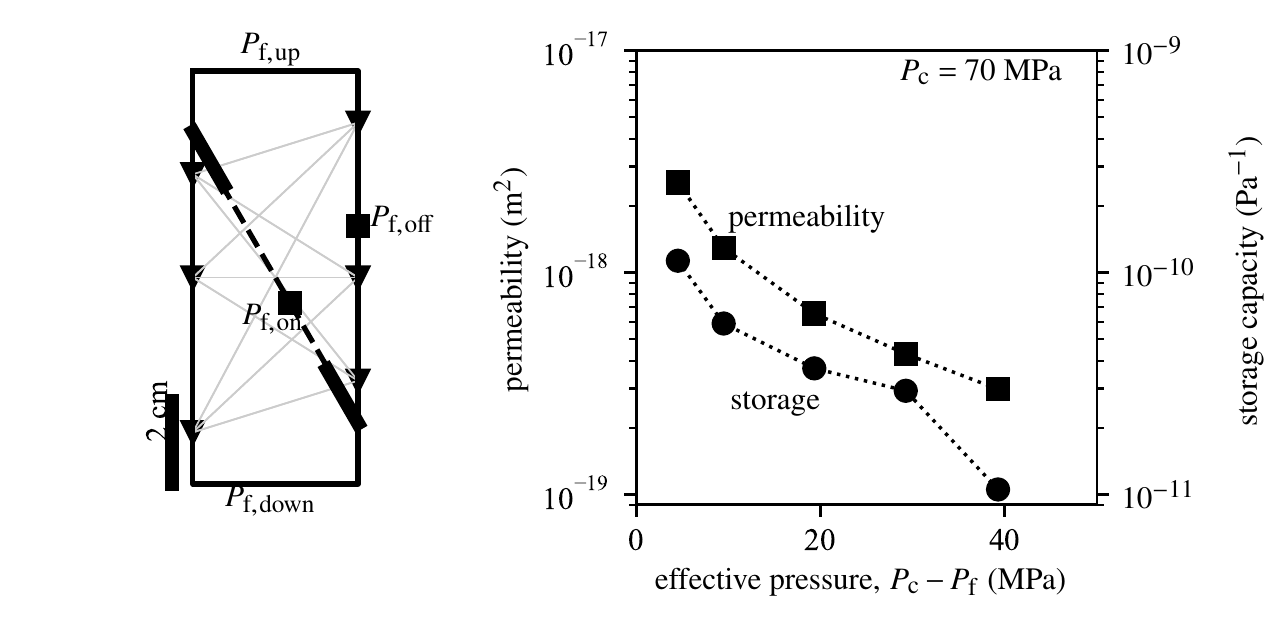}
  \caption{(left) Sample geometry in cross-section, showing the position and length of the 30$^\circ$ notches (thick black lines), the position of fluid pressure transducers (black squares), the position of the piezoelectric transducers (inverted triangles) and the ray paths between them (grey lines). (right) Evolution of permeability (squares) and storage capacity (circles) as a function of the Terzaghi effective pressure under hydrostatic conditions at $P_\mathrm{c}=70$~MPa and step-wise increases in pore pressure.}
  \label{fig:sample}
\end{figure}

The deformation experiments were conducted on thermally-cracked, notched Westerly granite samples equipped with two to four fluid pressure transducers and, in one instance, a set of piezoelectric transducers capable of measuring ultrasonic wave velocities. Cylindrical cores of 40~mm in diameter and 100~mm in length were machined and their faces ground parallel. Two aligned notches of 1.5~mm in width and 17~mm in length were cut at an orientation of 30$^\circ$ from the axis of the cylinders on opposite sides of the sample (Figure \ref{fig:sample}), in order to favor the propagation of rupture along a predictable plane. The samples were then thermally cracked at 600$^\circ$C during 5~hours (using a ramp of 4$^\circ$C per hour), which ensured the formation of a permeable microcrack network \citep[e.g.,][]{wang13}. The notches were filled with 1.5~mm thick Teflon sheets, and the samples were jacketed in a 3~mm thick Viton sleeve. Experiments were conducted on two samples. One sample was equipped with 3 pairs of piezoelectric transducers placed in a plane perpendicular to the prospective fault plane, and two differential fluid pressure transducers were placed on the material, one along the prospective fault plane, and one perpendicular to it (see Figure \ref{fig:sample}, left panel). The other sample was equipped with four differential pressure transducers, three along the prospective fault plane and one across it (Figure S1).

The differential fluid pressure transducers are made of a hollow steel insert facing the sample surface on one side, and closed on the other side by a steel cap sealed by an O-ring (Figure S2). The steel cap is machined on the inside to form a penny-shaped cavity which is connected to the pore space of the rock through the insert, and sealed from the confining medium. The exterior face of the steel cap is equipped with a diaphragm strain gauge sensitive to the elastic distortion of the cap in response to variations of the differential pressure between the confining medium and the pore fluid. The output of the diaphragm strain gauge is amplified with a dynamic bridge amplifier. The fluid pressure transducers were calibrated in situ by changing the uniform pore pressure in the sample while keeping the confining pressure constant (Figure S3). A linear output was observed for all transducers.

Jacketed samples were placed inside the pressure vessel of the UCL triaxial Rock Physics Ensemble \citep{eccles05}. Prior to deformation and fracture, the poroelastic and hydraulic properties of the sample were characterised during a series of hydrostatic loading and unloading cycles (see Figures S4, S5, S6 and S7 for details). The hydraulic properties of the undeformed sample (Figures \ref{fig:sample}, right panel) show a strong dependence on effective pressure, typical of cracked granite \citep[e.g.,][]{wang13}.

The sample equipped with two pore pressure transducers and piezoelectric transducers was tested at a confining pressure $P_\mathrm{c}=70$~MPa and an initially uniform pore pressure $P_\mathrm{f}=30$~MPa. The additional test on the sample equipped with four pore pressure transducers was conducted at $P_\mathrm{c}=60$~MPa and $P_\mathrm{f}=20$~MPa. The deformation tests were performed at a constant axial shortening rate of $10^{-6}$~s$^{-1}$, and a constant pore pressure (20 or 30~MPa) was maintained at both ends of the sample (downstream and upstream). The volume in the pore fluid intensifier was recorded to give access to the finite pore volume change in the samples when pore pressure was equilibrated throughout the pore and tubing network. All mechanical data, including outputs of fluid pressure transducers, were recorded at a nominal rate of 1~Hz, and the sampling rate was increased to 5~Hz during rupture and slip events. 

\section{Results}

\subsection{Fluid pressure drop during rupture}

\begin{figure}
  \centering
  \includegraphics{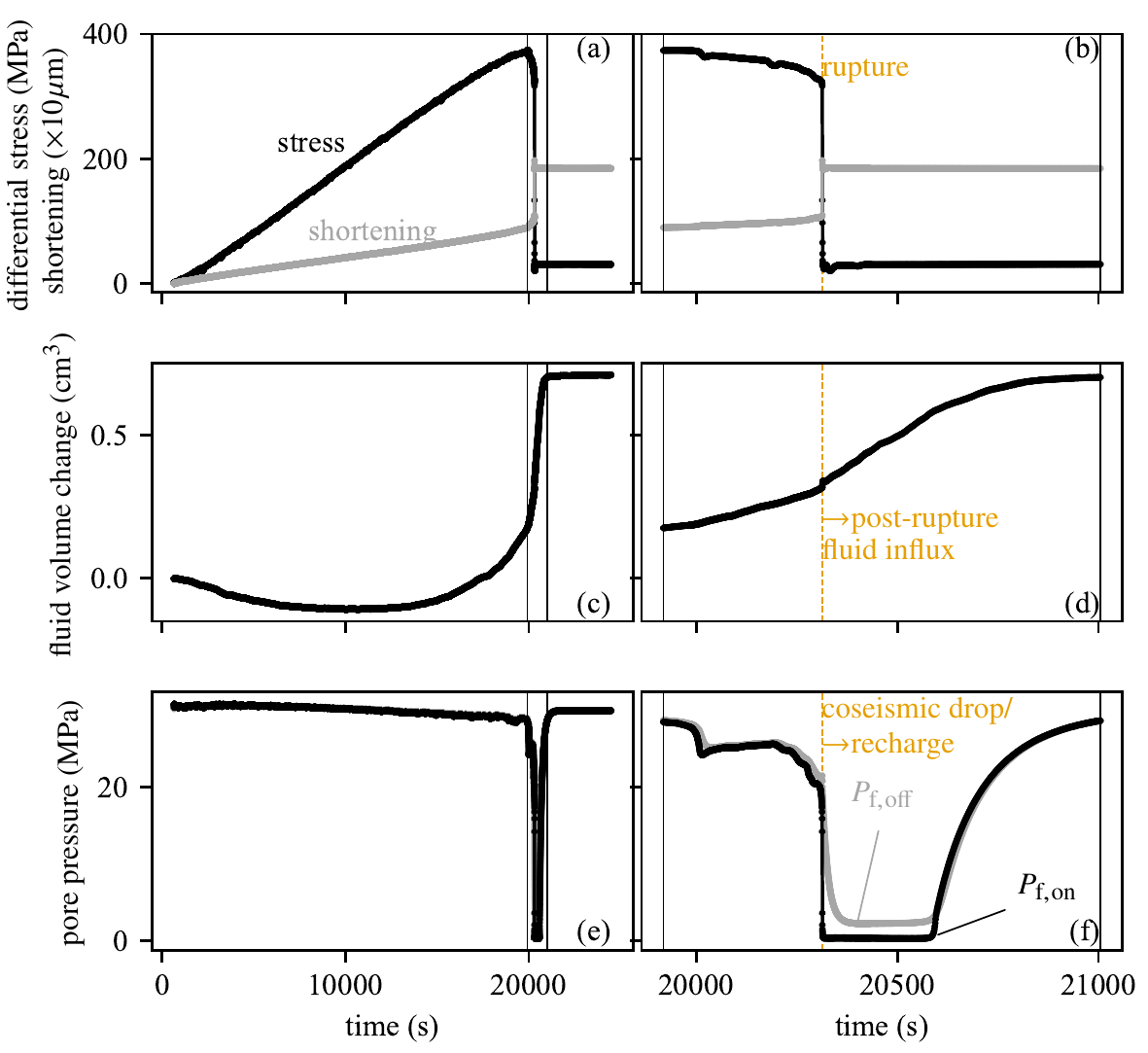}
  \caption{Test conducted at $P_\mathrm{c}=70$~MPa and $P_\mathrm{f}=30$~MPa. Differential stress and axial shortening (a,b), fluid volume change (c,d) and on- and off-fault pore pressure (e,f) as a function of time during the triaxial rupture experiment. Panels (b,d,f) correspond to the time period marked between the thin vertical lines in panels (a,c,e).}
  \label{fig:rupture}
\end{figure}

During deformation in the test conducted at $P_\mathrm{c}=70$~MPa and $P_\mathrm{f}=30$~MPa, the differential stress initially increased linearly with increasing axial strain (Figure \ref{fig:rupture}a), and the pore volume initially decreased, reached a minimum and then increased with increasing deformation (Figure \ref{fig:rupture}c). Concomitantly, the pore pressure measured inside the sample remained almost constant (within a few megapascals) up until the peak stress (Figure \ref{fig:rupture}e). In the few minutes prior to macroscopic failure of the sample, the applied stress reached a peak, decreased progressively and underwent a small sudden drop (Figure \ref{fig:rupture}b), accompanied by a small drop in both on- and off-fault pore pressure ($P_\mathrm{f,on}$ and $P_\mathrm{f,off}$, respectively) (Figure \ref{fig:rupture}f). With increasing deformation, the applied stress decreased further, accompanied by accelerated deformation and dilatancy. Together with this acceleration, the internal fluid pressure measured on the fault trace $P_\mathrm{f,on}$ started decreasing rapidly, and suddenly dropped to zero (within the accuracy of the transducer calibration) during the macroscopic failure of the sample (Figure \ref{fig:rupture}f). The pore pressure measured off the fault trace, $P_\mathrm{f,off}$, followed a similar pattern but after some delay, and its drop was more progressive and toward a small but nonzero value $P_\mathrm{f,off}\approx2.3$~MPa. This overall pattern was reproduced in the additional test performed at $P_\mathrm{c}=60$~MPa and $P_\mathrm{f}=20$~MPa (Figure S8).


After rupture, $P_\mathrm{f,on}$ remained around zero and $P_\mathrm{f,off}$ remained at $2.3$~MPa for around $250$~s, while the pore fluid volume (as measured externally from the pore pressure intensifier volume change) kept increasing almost linearly with time. In the final stage, $P_\mathrm{f,off}$ started increasing, followed by $P_\mathrm{f,on}$, asymptotically reaching the set pressure $P_\mathrm{f,up}=P_\mathrm{f,down}=30$~MPa. Considering a pore pressure difference of the order of $2$~MPa between the off- and on-fault locations, measurement of fluid flow rate into the fault zone during the post-rupture fluid recharge yields an off-fault permeability of around $4\times10^{-18}$~m$^2$, which is comparable to the permeability of the undeformed material at near-zero effective pressure (Figure \ref{fig:sample}).

\subsection{Wave speed variations during rupture}

\begin{figure}
  \centering
  \includegraphics{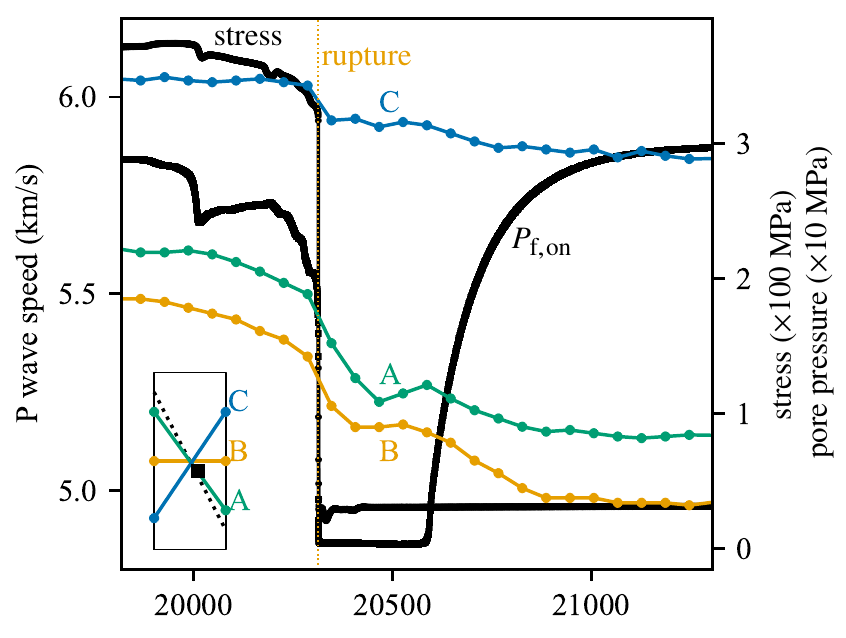}
  \caption{Test conducted at $P_\mathrm{c}=70$~MPa and $P_\mathrm{f}=30$~MPa. Evolution of differential stress, on-fault pore pressure, and P wave speed along (path A, green) and across (paths B, orange, and C, blue) the fault during rupture.}
  \label{fig:vprupture}
\end{figure}

The P wave speed of the undeformed sample was initially uniform and isotropic, equal to around $6$~km/s. Prior to the main rupture event, the P wave speed measured along the fault trace and along the horizontal (perpendicular to the compression axis) decreased progressively from the intact value down to around $5.5$~km~s$^{-1}$ and $5.3$~km~s$^{-1}$, respectively (Figure \ref{fig:vprupture}, path A and B). By contrast, the P wave speed along a subvertical direction across the fault trace did not shown any significant decrease before rupture (Figure \ref{fig:vprupture}, path C). While the main rupture event itself was very sudden, the P wave speed measured along the fault and along the horizontal axis (paths A and B) decreased progressively and stabilised while the on-fault fluid pressure remained zero. During the fluid pressure recharge towards $P_\mathrm{f,on}\rightarrow30$~MPa, the P wave speed (paths A and B) decreased again progressively. The P wave speed averaged across the fault in a subvertical orientation (path C) exhibited the same features but the amplitude of the change was much lower (only around $100$~m~s$^{-1}$ just after rupture) and more gradual, highlighting the highly localised nature of the variations in wave speeds.

In summary, the wave speed measurements show a clear two-step decrease along paths A and B:  a first large drop, followed by a plateau and then a smaller gradual decrease. Along path C (diagonal orientation across the fault), only a small, gradual decrease is observed after rupture.


\subsection{Fluid pressure drop during subsequent slip events}

\begin{figure}
  \centering
  \includegraphics{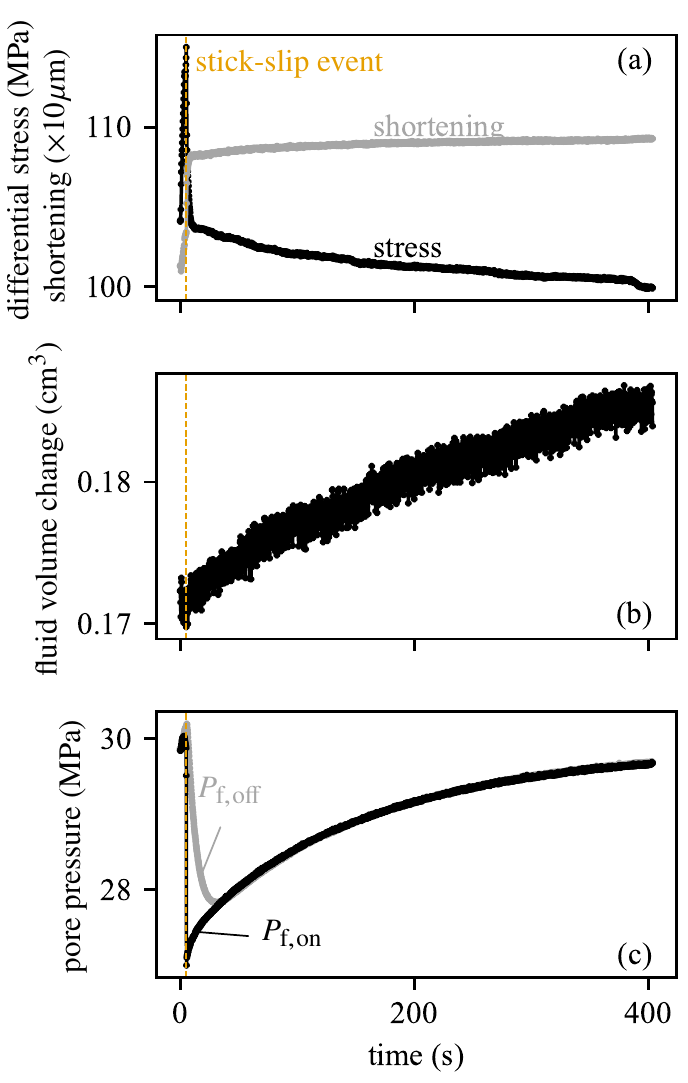}
  \caption{Test conducted at $P_\mathrm{c}=70$~MPa and $P_\mathrm{f}=30$~MPa. Differential stress and axial shortening (a), fluid volume change (b) and on- and off-fault pore pressure (c) as a function of time during a stick-slip event.}
  \label{fig:stickslip}
\end{figure}

After the main rupture, further axial shortening resulted in a series of dynamic stick-slip events along the newly-created fault. Prior to stick-slip, the differential stress increased elastically (Figure \ref{fig:stickslip}a), which was accompanied by a small increase in on- and off-fault fluid pressure (Figure \ref{fig:stickslip}c). Stick-slip was marked by a sudden stress drop and a step in axial shortening, and systematically accompanied by a sudden drop in on-fault fluid pressure, typically of a few megapascals in amplitude (Figure \ref{fig:stickslip}c), followed by a slow recovery. Compared to the on-fault fluid pressure, the off-fault fluid pressure decreased more progressively and the net drop was smaller. The fluid pressure recovery occurred at a similar rate for both on- and off-fault locations, and was accompanied by an influx of fluid (as measured by the total pore volume change, Figure \ref{fig:stickslip}b). No significant wave velocity change was measured prior to or after stick-slip events.

\begin{figure}
  \centering
  \includegraphics{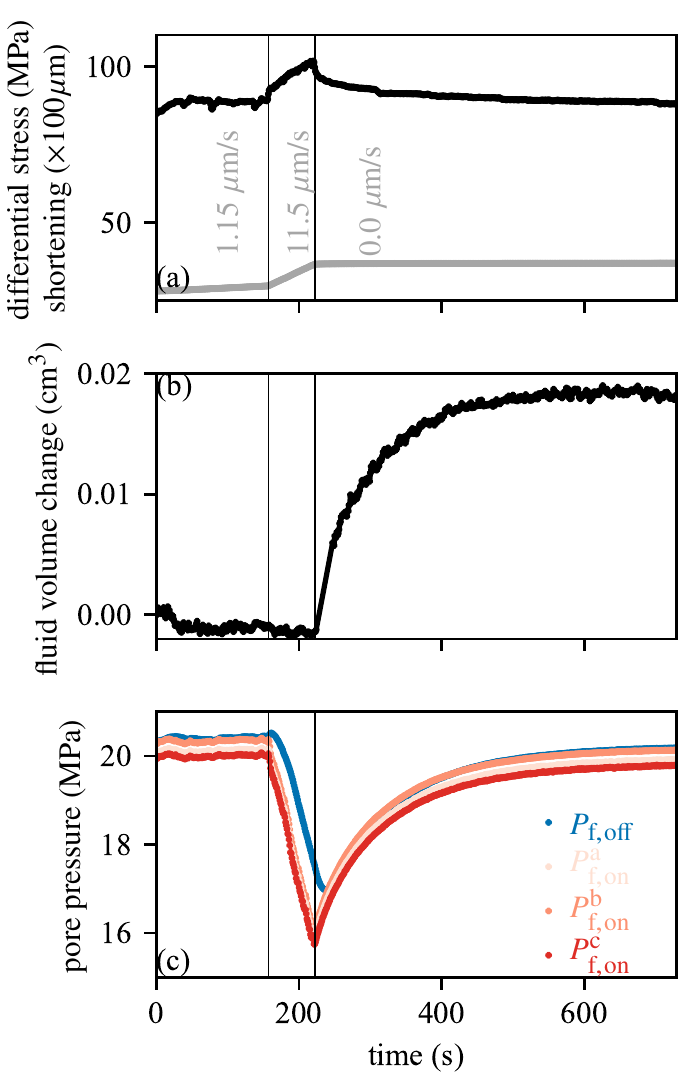}
  \caption{Test conducted at $P_\mathrm{c}=60$~MPa and $P_\mathrm{f}=20$~MPa. See Figure S1 for sample geometry and labelling of pore pressure transducers. Differential stress and axial shortening (a), fluid volume change (b) and on- (red) and off-fault (blue) pore pressure (c) as a function of time during a slip-rate step from $1.15$ to $11.5$~$\mu$m/s.}
  \label{fig:slip}
\end{figure}

In addition to stick-slip events, stable slip was also observed. A series of slip rate steps were conducted at a slip rate of $11.5$~$\mu$m/s for durations ranging from a few seconds to tens of seconds. The on-fault fluid pressure decreased continuously with on-going slip, while the intensifier volume remained constant (Figure \ref{fig:slip}). The fluid pressure measured off-fault also decreased, but with a significant delay (Figure \ref{fig:slip}c, blue curve), indicating that the pore volume source was localised on the fault. The lack of fluid flow from the intensifier observed during rapid slip despite the significant pore pressure drop in the sample indicates that the fault zone was essentially undrained; in addition, poroelastic compaction of the intact regions of the sample, near the drained ends, could also contribute to compensate the induced inflow as differential stress increases. When slip was stopped, the pore pressure gradually recovered and the intensifier volume change indicates that a significant pore volume increase occurred. 


\begin{figure}
  \centering
  \includegraphics{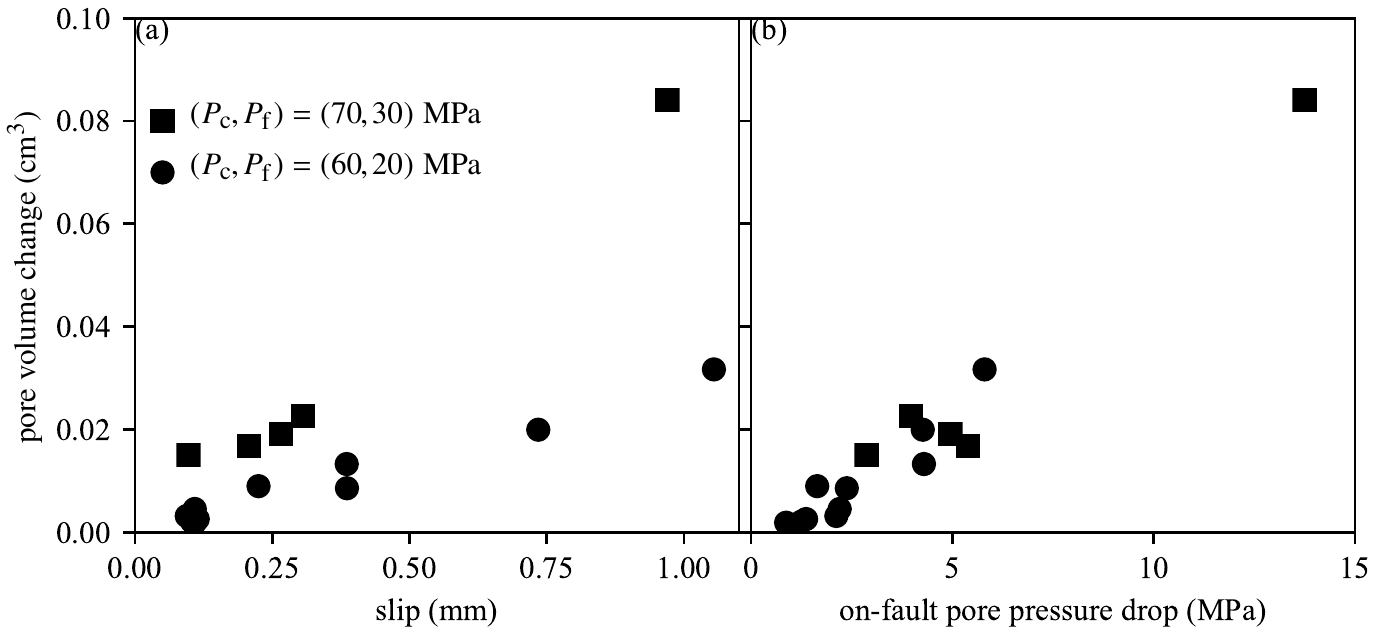}
  \caption{Pore volume change as a function of slip (a) and on-fault peak pore pressure drop (b) during all slip episodes (controlled at $11.5$~$\mu$m/s and stick-slip).}
  \label{fig:dpeventsall}
\end{figure}

The observation of a pore pressure drop during slip was systematic for slip rates above 10~$\mu$m/s. After each imposed slip rate episode, shortening was stopped until complete reequilibration of pore pressure throughout the sample. By recording the net variation in pore fluid intensifier volume after slip and pore pressure reequilibration, an estimate of the sample pore volume change incurred by fault slip is obtained (Figure \ref{fig:dpeventsall}a). For both tests, the pore volume change increased more or less linearly with increasing imposed slip. In the sample fractured at $P_\mathrm{c}=70$~MPa and nominal $P_\mathrm{f}=30$~MPa, the maximum recorded pore volume change was of around 0.08~cm$^3$ after 0.97~mm slip. By contrast, in the sample fractured at $P_\mathrm{c}=60$~MPa and $P_\mathrm{f}=20$~MPa, the largest recorded pore volume change was of 0.03~cm$^3$ at 1.04~mm slip. Using a representative, rather conservative fault width of 3~mm, the observed pore volume change translate to a maximum of $+1.12$\% and $+0.42$\% fault zone porosity increase after around 1~mm slip in the sample tested at $P_\mathrm{c}=70$ and $P_\mathrm{c}=60$~MPa, respectively. The difference in dilatancy rate per unit slip between the two samples tested indicates some variability in either fault geometry or internal structure that could be attributed to the slightly different pressure conditions tested,  but also to the natural variability in fault structures formed in experimentally faulted granite.

The peak pore pressure drop recorded during each slip episode increases approximately linearly with increasing pore volume change, in a trend common to both samples (Figure \ref{fig:dpeventsall}b). The pore pressure drop recorded by on-fault transducers is a lower bound for the actual pore pressure variation inside the fault, due to (1) the nonzero internal volume of the transducers, and (2) nonzero drainage from the pore space within the fault walls, notably during controlled slip episodes at $11.5$~$\mu$m/s. Despite those limitations, the overall trend of pore pressure drop as a function of net pore volume change appears quite robust. The variability observed during slip episodes associated with low pore volume change is likely due to the existence of a range of drainage conditions, from near-perfectly undrained events during stick-slip to partially drained events during fast but controlled slip. In all cases, stick-slip events were preceded by some stable, controlled slip, so it was not possible to completely separate the pore volume change associated with each phase. The approximately linear relationship between pore pressure change and pore volume change provides an estimate of the compressibility of the fault zone material, here of the order of $C_\mathrm{pp} = (1/\phi)\partial\phi/\partial P_\mathrm{f}\approx 2 - 4\times 10^{-8}$~Pa$^{-1}$, using a nominal fault porosity $\phi_0$ between 2 and 4\%.

\section{Microstructures}

\begin{figure}
  \centering
  \includegraphics{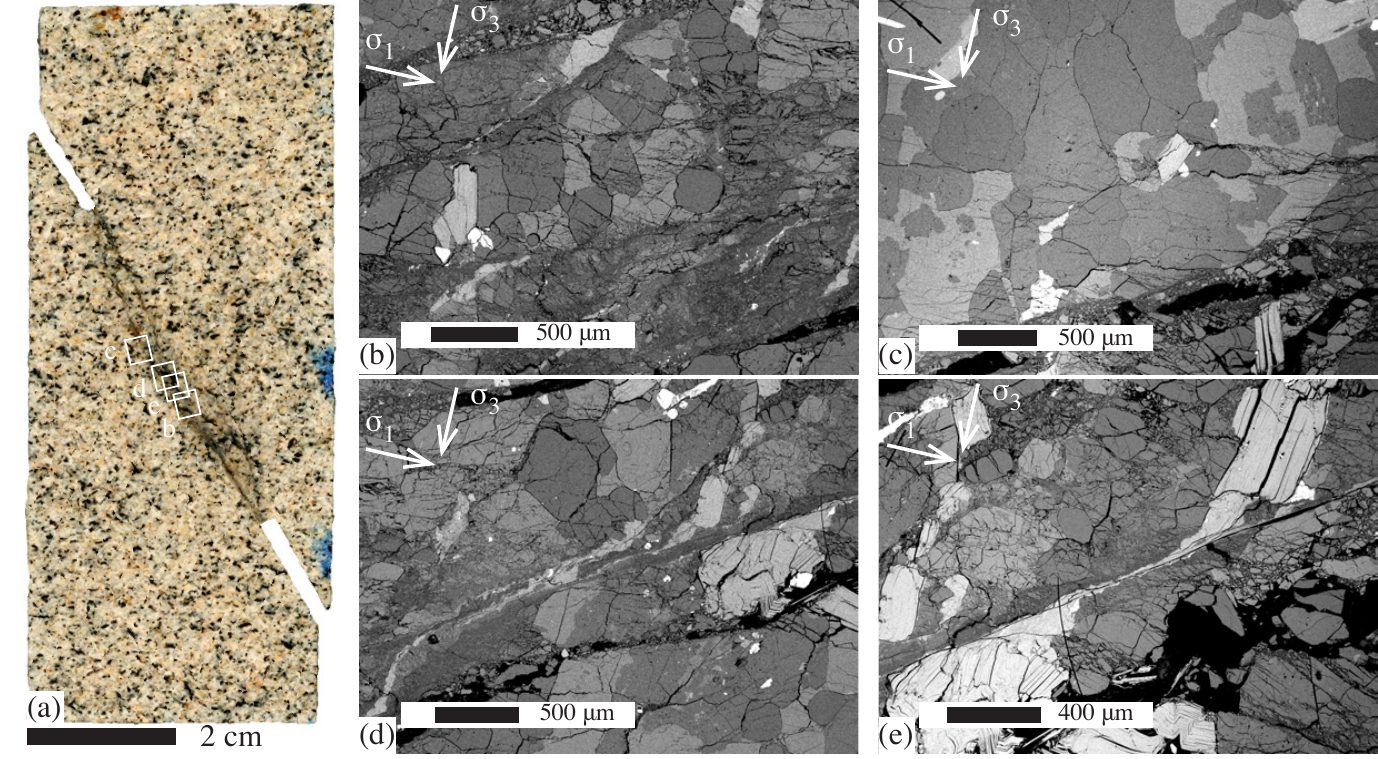}
  \caption{Fault zone structure in the sample tested at $P_\mathrm{c}=70$~MPa and $P_\mathrm{f}=30$~MPa. (a) Optical image of the whole sample after polishing (notches were filled with a Teflon sheet during the test itself). (b--e) SEM Backscattered electron images of the internal structure of the fault; the orientation of remotely applied principal stresses is indicated by the white arrows.}
  \label{fig:microstr1}
\end{figure}

A clear fault zone was formed and slipped in both tested samples. Detailed observations were performed of the fault structure formed during the test conducted at $P_\mathrm{c}=70$~MPa and and $P_\mathrm{f}=30$~MPa. After vacuum impregnation in epoxy resin, the sample was cut perpendicular to the fault zone and the entire surface was polished down to a 0.5~$\mu$m grit. At sample scale, the fault zone connects the two notches at the average angle of 30$^\circ$, but contains more than one strand and appears undulous (Figure \ref{fig:microstr1}a). The internal structure of the fault zone, observed with a Scanning Electron Microscope (SEM), consists
of a gouge zone delimited by two main boundary faults (Figure \ref{fig:microstr1}b).  The overall fault zone thickness where intense microfracturing and cataclasis is observed is of around 3~mm. The transition from the gouge zone to fault walls is sharp, and off-fault damage is present in the form of thin intra- and inter-granular tension cracks oriented along the compression direction located within 1 to 2~mm from the gouge boundary (Figure \ref{fig:microstr1}c). The internal structure of the gouge zone itself is complex. Multiple cataclastic shear zones oriented at 30$^\circ$ from the compression axis cut heavily microcracked lenticular pieces of the parent rock (Figure \ref{fig:microstr1}b). In addition, thin shear zones with intense grain size reduction are present in the central part of the gouge zone, with offsets of several millimetres highlighted by sheared mica grains (Figures \ref{fig:microstr1}d,e). The total slip accumulated across the fault was 5.2~mm, which is significantly larger than the largest of the observed offsets across any single thin cataclastic band, which indicates that several bands must have operated either sequentially or collectively to accommodate the total imposed slip.

\begin{figure}
  \centering
  \includegraphics{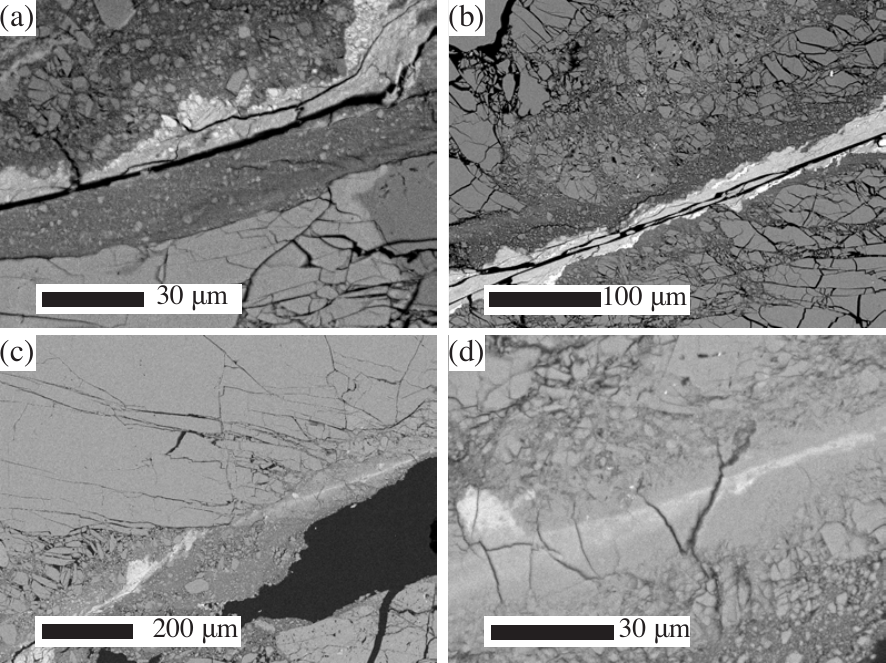}
  \caption{Fault zone structure in the sample tested at $P_\mathrm{c}=70$~MPa and $P_\mathrm{f}=30$~MPa. (a--d) SEM Backscattered electron images of the thin shear zones occurring within the fault gouge layer. In image (c), the large black region is an area where the gouge material was lost during sample preparation.}
  \label{fig:microstr2}
\end{figure}

The thin shear bands consist mostly of dense, finely comminuted material (Figure \ref{fig:microstr2}a), sometimes containing at their center some material derived from a sheared mica, with undulating boundaries and flow structures (Figure \ref{fig:microstr2}b). In places, the material forming the shear zone transitions from a dense, fine-grained granular aggregate to a completely nonporous continuous matrix containing traces of chemical heterogeneities aligned with the shear zone (Figure \ref{fig:microstr2}c,d). These textures are very similar to those reported by \citet{hayward17} in faulted sandstone, and could be interpreted as welded patches made of glass, potentially cooled from a frictional melt. More detailed analyses would be needed to confirm the potential melt origin of such textures, knowing that amorphous material can be produced without melting \citep[e.g.,][]{yund90,pec12}. In any case, such an extreme grain size reduction, compaction, and the presence of elongated flow structures indicate that strain was strongly localised along a set of internal shear zones after the main fault-forming rupture event. 

Overall, microstructural investigations indicate that the dilation responsible for pore pressure fluctuations during rupture and slip was likely localised within the fault zone itself, as evidenced by the heavily cracked, porous gouge layer formed inside the fault, and the relatively narrow damage zone around it. The overall waviness of the fault likely contributed significantly to slip-induced opening.






\section{Discussion}

\subsection{Fluid pressure drop and localised dilatancy during rupture}

The experimental data show a systematic drop in fluid pressure during the main rupture and each slip event, associated with localised dilatancy in the fault zone. These data provide direct evidence of the ``seismic suction pump'' model of \citet{sibson87}. During rupture, the measurements of on-fault pore pressure show very rapid drops to near-zero values, while off-fault measurements are more gradual and do not reach zero. These results can be explained by localised dilatancy within the fault zone, which depressurises directly the fluid inside the fault, and induces a diffusive fluid flux that makes the pressure drop gradually propagate into off-fault regions.

The on-fault fluid pressure remained stable at near-zero values during a few minutes after rupture (Figures \ref{fig:rupture}f, S8f). If a constant fluid compressibility was assumed, we would expect an immediate pore pressure recharge after the drop (similar to that observed immediately after stick-slip events, where the pore pressure drop was relatively modest, see Figure \ref{fig:stickslip}c). This discrepancy is resolved by considering that the pore fluid either degasses or vaporises due to rapid isothermal decompression, producing an apparent increase in fluid compressibility, therefore stabilising the fluid pressure in the fault, and buffering the fluid pressure off the fault.

In the sample tested at $P_\mathrm{c}=70$~MPa and $P_\mathrm{f}=30$~MPa, the total pore volume change measured after complete post-failure reequilibration was $0.37$~cm$^3$, which corresponds to a porosity change of $+5$\% in a $3$~mm-wide fault zone. Under undrained, isothermal conditions, the fluid mass conservation leads to a pressure change expressed as
\begin{linenomath}\begin{equation} \label{eq:dp}
  \Delta p =  -\ln(\phi\rho(p)/\phi_0\rho_0)/{C_\mathrm{pp}},
\end{equation}\end{linenomath}

where ${C_{\mathrm{pp}}}$ is the pore space compressibility, $\phi$ and $\phi_0$ are the final and initial porosity, $\rho_0$ is the initial fluid density and $\rho(p)$ is the fluid density at pressure $p$. Neglecting the compressibility of the solid grains of the rock and using an initial porosity of $\phi_0=2$\%, an upper bound estimate for ${C}_{\mathrm{pp}}$ is $5\times10^{-9}$~Pa$^{-1}$ (estimated as the ratio of storage capacity and porosity, see Figures \ref{fig:sample} and S5). Starting from $30$~MPa and using pressure-dependent variations in fluid density \citep{junglas09}, the vaporisation pressure ($\approx3$~kPa at room temperature) is reached after a porosity increase of only $0.3$\%. Even considering a pore space compressibility increased by a factor $10$ to simulate the effect of damage \citep[e.g.,][]{noda09,brantut18c}, the vaporisation pressure is reached after a porosity increase of around $3$\%. Therefore, the pore fluid could be at least partially vaporised due to the fault zone dilatancy during rupture. It is unclear how much dissolved gas was present in the pore fluid prior to rupture events, but degassing of dissolved air is expected to occur during decompression, which could also contribute significantly to increasing the apparent compressibility of the pore fluid and stabilising the pore pressure after rupture.

In the experiments, shear failure of initially intact (notched) rock samples produced spontaneously a non-planar fault geometry at centimetre scale (Figure \ref{fig:microstr1}a), which allowed us to capture multiple dilatancy mechanisms: initial coalescence of microcracks to form the through-going fault, granular packing variations within the gouge, microfracturing of wall rock (off-fault) and of cohesive lenses of material embedded in the gouge (Figure \ref{fig:microstr1},b,c), and opening of dilational jogs. Taken together, these dilatant processes resulted in dramatic pore pressure drops during rupture. The non-planar geometry of natural faults, with the presence of local dilational jogs, has been considered as a key factor controlling fluid pressure variations and redistribution along faults \citep{sibson87,sibson94}. The experimental data therefore provide direct evidence supporting the conclusions drawn from field observations that coseismic dilatancy can locally depressurise fluids and induce phase separation, mineral precipitation and gold deposition \citep[e.g.,][]{sibson87, wilkinson96, cox99, wilkinson01, weatherley13}.

Here, only dilatancy is observed with increasing slip, whereas a number of experimental studies on synthetic gouges report gouge compaction with increasing slip \citep[e.g.,][]{morrow89,scuderi15}. Gouge compaction during stick-slip events is often attributed to granular rearrangements within the gouge, while gouge dilation tends to be observed in the phase preceding slip. Such a behaviour can be thought of as an oscillation around a well-defined ``steady-state'' in terms of gouge porosity (equivalent to critical state in the terminology of soil mechanics). Here, the complex geometry of the fault and the spontaneous generation of gouge from the intact rock imply that the fault zone porosity is still far from steady-state, and new gouge material and void opening keep being generated as slip proceeds. The experiments are limited to small slip distances (a few mm), but one expects that the rate of dilatancy should decrease with increasing slip as the fault matures, and the purely dilatant behaviour of the experimental fault might transition to a more complex behaviour dominated by internal gouge particle rearrangements.


\subsection{Seismic signature of dilatancy during rupture}

In the experiments, the signal observed in seismic wave speeds was quite complex (Figure \ref{fig:vprupture}): while some areas and orientations were severely impacted by the main rupture and a clear signal could be attributed to pore pressure recharge (see the second drop in P wave speed along paths A and B), such observations are not ubiquitous and depend on the details of the ray paths geometry. This complexity reflects the localisation of microcrack damage during rupture \citep{aben19}, which in turn impacts the localisation of dilatancy and pore pressure variations. Dilatant microcracking has two effects on wave velocities: (1) a direct effect, whereby microcracks tend to decrease elastic wave speeds (preferably in orientations perpendicular to crack faces, see for instance \citet{sayers95}), and (2) an indirect effect, whereby increases in pore pressure tend to open existing or newly formed microcracks through an decrease in effective mean stress. The direct effect is very strong and localised, leading to the first observed drop in P wave speed along paths A and B. The indirect effect is of second order, but leads to clear variations as evidenced by the second, more gradual decrease in wave speed observed in the experiment that is associated with an increase in pore pressure (Figure \ref{fig:vprupture}). The slight increase at the end of the pore pressure plateau (paths A and B, Figure \ref{fig:vprupture}) is consistent with partial vaporisation or degassing: in partially saturated rocks, P wave speed increases significantly as full saturation is approached \citep{bourbie86}.

While a sufficient level of detail can be achieved in laboratory observations to provide a clear interpretation of the results, evidences for pore pressure drops would be challenging to detect seismologically in a systematic way in nature. Seismological data consistently indicate that wave speeds decrease around fault zones after earthquakes \citep[e.g.,][]{tadokoro02,schaff04,li06,froment14}, which has been attributed to damage generation or reactivation. However, such observations typically represent a spatial and temporal average of the changes within a volume surrounding the fault. The effect of a pore pressure drop on wave speeds is of second order, and is linked to the positive correlation between wave speeds and effective pressure in fractured media. The effect of pore pressure change on wave speeds is not only small in comparison to the direct effect of microcrack damage, but is also strongly localised in space and transient in time. Therefore, wave speed monitoring would require a high spatio-temporal resolution together with a high precision to be interpreted unambiguously.

\subsection{Coseismic dilatancy hardening and impact on weakening mechanisms}

Dilatancy is shown here to be the dominant process driving pore pressure change during the early stages of slip along a fresh fault with realistic roughness. Dilatancy effectively resets the fluid pressure to lower values at the onset of seismic slip, and tends to counteract the thermal pressurisation process as slip proceeds \citep{rice06}. The two samples tested showed dramatic pore pressure drops during rupture and fault formation, and during subsequent slip events. The dataset being quite limited (only one rock type was tested, and deformation was limited to small slip distances), one should refrain from producing overly detailed quantitative extrapolations to crustal faults. Nevertheless, the order of magnitude of the pore pressure drops is robust and reproducible, and the experimental data provide unique constrains on the key parameters governing the effect of dilatancy on pore pressure variations during faulting.

Microstructural observations indicate that the sample pore volume change was localised in the fault zone. Therefore, the evolution of net pore volume change as a function of slip (as measured by taking the difference between intensifier volume before and after slip) can be used to make elementary predictions of the expected pore pressure change during earthquakes. The pore volume change appears to be roughly a linear function of the accumulated slip (Figure \ref{fig:dpeventsall}a). The rate of pore volume change per increment of slip differs quite significantly (by a factor of around 2.5) between the two experiments, but remains of a similar order of magnitude. Such a variability likely reflects slight variations in fault geometry due to the spontaneous rupture process, as well as the use of a different combination of confining and nominal fluid pressure between the two tests. Regardless of such variations, a linear increase in fault opening with increasing slip can be explained by the rough fault profile, with overriding asperities generating net opening as slip proceeds. Observations are limited since only a few millimetres of slip could be imposed during the experiments, but it is unlikely that fault opening can keep increasing linearly as very large slip is accumulated. Fault wear rates decrease dramatically after the first few centimetres of slip \citep[e.g.,][Chap. 2]{scholz19}. In addition, as the fault matures, fault wear (and the associated microcracking and dilatancy) is expected to reduce the roughness along strike and thicken the gouge zone \citep[][Chap. 3]{scholz19}. Therefore, extrapolation of the experimental dilatancy data to a wider range of slip distances requires a transition from large to small dilation rate beyond some critical slip distance.


A simple, natural quantitative description of the change in porosity during slip is therefore given by 
\begin{linenomath}\begin{equation} \label{eq:dphi}
  \phi(\delta) = \phi_0 + (\phi_\mathrm{max}-\phi_0)(1-e^{-\delta/\delta_\mathrm{D}}),
\end{equation}\end{linenomath}
where $\phi_0$ is the initial porosity of the fault, $\phi_\mathrm{max}$ is the asymptotic porosity at large slip, $\delta$ is slip and $\delta_\mathrm{D}$ is a characteristic slip displacement over which dilatancy occurs. A realistic value for the maximum porosity $\phi_\mathrm{max}$ is that of a random packing of spheres, which is around 35\%. Considering a fault width of 3~mm, the experimental measurements of fluid volume change after each slip event (Figure \ref{fig:dpeventsall}a) can be converted into fault zone porosity changes ranging from $+0.42$\% to $+1.12$\% at around 1~mm slip. Values of $\delta_\mathrm{D}=0.08$ and $0.03$~m, respectively, are in agreement with these observations. The measurements of pore pressure drop associated with each slip event (Figure \ref{fig:dpeventsall}b) provide a constrain on the pore space compressibility ${C}_\mathrm{pp}\approx 2-4\times 10^{-8}$~Pa$^{-1}$. 

\begin{figure}
  \centering
  \includegraphics{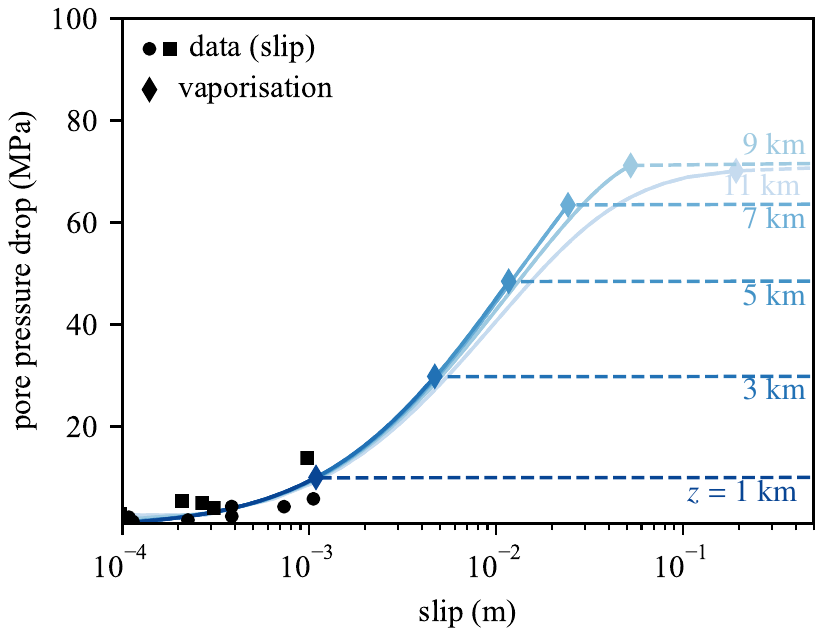}
  \caption{Isothermal, undrained pore pressure drop as a function of slip for a range of depths. Black circles and squares show laboratory data obtained during slip in the tests conducted at $(P_\mathrm{c},P_\mathrm{f})=(70,30)$ and $(P_\mathrm{c},P_\mathrm{f})=(60,20)$, respectively. Diamonds indicate points where fluid vaporisation occurs. An initial porosity of $2$\% was used. At each depth, the initial pore pressure is assumed hydrostatic, and the initial temperature is computed using a $40^\circ$C~km$^{-1}$ geotherm.}
  \label{fig:pdrop}
\end{figure}


Combining the porosity model (Equation \ref{eq:dphi}) with the governing equation for fluid pressure change (Equation \ref{eq:dp}), the isothermal, undrained pore pressure drop is computed for a range of slip distances and depths across the seismogenic crust (Figure \ref{fig:pdrop}, where average parameter values of $\delta_\mathrm{D}=0.05$~m, $\phi_0=2$\%, $\phi_\mathrm{max}=35$\% and $C_\mathrm{pp}=3\times 10^{-8}$~Pa$^{-1}$ have been assumed). Under these conditions, total fluid pressure drops (i.e., down to degassing or vapor pressure) are predicted at the early stages of slip (typically less than 1~cm) throughout the upper 9~km of the crust. Considering the variability and limited accuracy of laboratory-derived parameters, and the simplicity of the proposed dilatancy model, the numerical results given in Figure \ref{fig:pdrop} should be viewed only as order-of-magnitude estimates. Significant deviations from these estimates are expected depending on the degree of fault core consolidation or healing (as observed experimentally, intact rocks experience dramatically more dilation than preexisting faults), as well as from local variations in fault geometry (e.g., more dilation expected in dilational jogs). While a systematic experimental characterisation of these effects is currently missing, the present data indicate that on-fault dilation contributes significantly to coseismic pore pressure changes, which potentially impacts dynamic weakening mechanisms such as thermal pressurisation.




Pore pressure drops due to dilatancy are unlikely to be overcome or even balanced by shear heating effects: the thermal pressurisation factor, which relates the temperature rise to the pore pressure rise, is expected to be very small in freshly faulted rocks (as in our experiments), because the pore space compressibility is much larger than in intact rocks. Using a representative value of ${C_\mathrm{pp}}=3\times 10^{-8}$~Pa$^{-1}$, a realistic estimate for the thermal pressurisation factor is $0.03$~MPa~K$^{-1}$ \citep{brantut16b}, so that the slip weakening distance associated with thermal pressurisation is of the order of $0.12$~m, which is significantly larger than the slip required to induce complete pore pressure drops in the upper part of the crust (Figure \ref{fig:pdrop}).


The slip-dependent dilatancy model proposed here (Equation \ref{eq:dphi}) based on friction tests along freshly fractured rocks differs from the rate-and-state dependent model of \citet{segall95}, which is based on quasistatic synthetic gouge experiments from \citet{marone90}. While the \citet{segall95} model has been widely used in the context of earthquake nucleation and slow slip \citep[e.g.,][]{segall10}, including couplings with thermal pressurisation \citep{segall06,segall12}, it has been unclear whether it could be used to simulate rapid seismic slip. In that model, fault gouge porosity variations are linked to slip rate in the form \citep{segall95}
\begin{linenomath}\begin{align}
  \frac{d\phi}{dt} &= -\frac{v}{d_\mathrm{c}}(\phi-\phi_\mathrm{ss}),\\
  \phi_\mathrm{ss}  &= \phi_0 + \epsilon\ln(v/v^*),
\end{align}\end{linenomath}
where $v$ is the slip rate, $v^*$ is a reference (slow) slip rate, $d_\mathrm{c}$ is a slip-weakening distance (of the order of 10~$\mu$m), and $\epsilon$ is a dilation parameter of the order of $10^{-4}$ as per \citet{marone90}. Following a step increase in slip rate from $v^*$ to a larger value $V$, for instance to simulate a sudden slip episode such as stick-slip, the porosity evolution is of the form 
\begin{linenomath}\begin{equation}
  \phi(\delta) = \phi_0 + (\epsilon\ln(V/v^*))(1-e^{-\delta/d_\mathrm{c}}),
\end{equation}\end{linenomath}
which is similar to our simple approximation (equation \ref{eq:dphi}). However, the order of magnitude of the parameters is vastly different. The critical slip distance $d_\mathrm{c}$ is approximately 1000 times smaller than our inferred $\delta_\mathrm{D}$, and for an increase in slip rate from, say, 1~$\mu$m/s to 1~m/s, the maximum porosity increase would only be of +0.14\%, far less than reported in our experiments on freshly fractured rock. The discrepancy between the two datasets and models can be explained by the rough fault geometry of the fresh faults, which appears to dominate their dilatant response.

While further analysis is required to investigate fully how dilatancy and thermal pressurisation are coupled during seismic slip, the present laboratory experiments clearly demonstrate the possibility that rupture induces pore pressure drops in crystalline rocks. These results highlight that the occurrence of dilatancy is likely the major difference between faults hosted in intact or healed crystalline rocks and mature faults containing fine-grained gouge and clay minerals. In the former, pseudotachylytes are not uncommon \citep{sibson06} and at depths of 10 to 20~km in the crust, where temperature is elevated and fluid-rock interactions are relatively fast, post-seismic fault sealing should rapidly cement fault rocks, so that fault reactivation during subsequent earthquakes requires re-fracturing of the fault core. In regions where cementation and fault sealing are fast compared to interseismic periods, large dilatancy effects are expected during earthquake propagation. In mature faults and at shallower depths, pseudotachylytes are rare and thin slip zones with clay-bearing gouges might allow very little overconsolidation and negligible dilatancy. There, thermal pressurisation might be a very effective mechanism \citep{rice06}. The role of dilatancy appears to be a key test to determine the dominant weakening mechanism in earthquakes: while seismological data are consistent with thermal pressurisation across wide range of magnitudes \citep{viesca15}, other mechanisms with similar seismological signatures are likely operating \citep[e.g., flash heating or melting,][]{nielsen08,ditoro11,violay15,brantut17b}. Our laboratory measurements of pore pressure drop on newly formed faults with realistic geometries and internal structures constitute a first clear observation that thermal pressurisation might be less ubiquitous than previously thought.



\section{Conclusions}

Direct pore pressure measurements show that during experimental shear faulting of granite, localised fault dilatancy is large enough to decompress the saturating pore fluid from several tens of megapascals down to vapor pressure. Slip on the newly formed faults is also accompanied with net dilation, and, for high enough slip rates, pore pressure drops. The roughness of the spontaneously formed faults plays a key role in producing such strong dilatant effects, which are orders of magnitude stronger than predicted by models based on gouge deformation in planar faults. The laboratory observations presented here provide a quantitative, direct evidence of the ``seismic suction pump'' concept developed by \citet{sibson87}. The large fault dilatancy occurring during faulting of intact rocks and slip on fresh faults indicate that dynamic weakening by thermal pressurisation is unlikely to be dominant in those materials, and that other weakening mechanisms must play a role. 

\section*{Acknowledgments}

Neil Hughes provided technical support for the manufacture of the fluid pressure transducers. SEM analysis was performed with technical assistance from James Davy. Discussions with Frans Aben, Dmitry Garagash, Nadia Lapusta, Philip Meredith and Alexandre Schubnel contributed to shape this project. Comments from two anonymous reviewers on a previous draft helped with the discussion of the results. Comments from Ze'ev Reches and Terry Tullis helped clarifying the manuscript. Financial support from the UK Natural Environment Research Council (grants NE/K009656/1, NE/M016471/1 and NE/S000852/1) and from the European Research Council under the European Union's Horizon 2020 research and innovation programme (project RockDEaF, grant agreement \#804685) is acknowledged.



\end{document}